\def\year{2016}\relax
\documentclass[letterpaper]{article}
\usepackage[final]{nips_dlps_2017}
\usepackage{times}
\usepackage{courier}

\usepackage{graphicx} 
\usepackage{caption}

\usepackage{natbib}
\usepackage{amsmath}

\usepackage{algorithm}
\usepackage{algorithmic}

\usepackage{color}
\begin{document}

\title{Deep Learning Reconstruction of Ultra-Short Pulses}
\author{Tom Zahavy$^{2*}$, Alex Dikopoltsev$1*$, Oren Cohen$^1$, Shie Mannor$^2$ and Mordechai Segev$^1$\\
 $^1$ Department of Physics and Solid State Institute \\
$^2$ Department of Electrical Engineering \\
$^*$ Both authors contributed equally to this manuscript \\
 The Technion - Israel Institute of Technology, Haifa 32000, Israel\\
}

\maketitle
\begin{abstract}
Ultra-short laser pulses with femtosecond to attosecond pulse duration are the shortest systematic events humans can create. Characterization (amplitude and phase) of these pulses is a key ingredient in ultrafast science, e.g., exploring chemical reactions and electronic phase transitions. Here, we propose and demonstrate, numerically and experimentally, the first deep neural network technique to reconstruct ultra-short optical pulses. We anticipate that this approach will extend the range of ultrashort laser pulses that can be characterized, e.g., enabling to diagnose very weak attosecond pulses.
\end{abstract}

\section{Introduction}

Ultra-short laser pulses are the shortest systematic events that can currently be created. They are typically used to measure physical and chemical phenomena. These pulses are currently being used in numerous applications including material and tissue processing, medical-imaging and research of light and matter \citep{zewail2000femtochemistry,delgado2011femtosecond,malinauskas2016ultrafast}. The term ultra-short pulses refer to light pulses whose width is below picoseconds ($10^{-12}$ sec), while the shortest recorded pulse today lasts for 43-attoseconds ($10^{-18}$ sec) \citep{baltuvska2003attosecond,Gaumnitz17}. In many applications, an event is measured in time by a shorter one (the ultra-short pulse) using optical techniques. The problem with these methods is that they require the measurement of the pulse itself, so one must use an even shorter one to measure it. But when a shorter event is not available is it still possible to measure it? 

To solve this chicken-and-egg paradox, reconstruction algorithms measure the pulse using replicas of itself. A common example are autocorrelation techniques including intensity autocorrelation \citep{kop1995phase}, spectral interferometry \citep{carquille1975new} and SPIDER \citep{iaconis1998spectral}. The problem with these techniques is that the recovery of the pulse phase from intensity measurements (known as the one-dimensional pulse phase retrieval problem) is not always injective. There are simply too many (usually infinitely many) pulses that correspond to a given spectrum.
Frequency-resolved optical gating (FROG, \cite{trebino1997measuring}), an established approach to solving this problem, relies on measuring convolutions (auto-correlation) between the pulse and time-shifted replicas of itself. 
Unlike standard techniques, FROG attempts to solve a two-dimensional phase-retrieval problem, which has only “trivial” ambiguities \citep{trebino2000frequency}.

The reconstruction of the pulse from the FROG measurement requires a recovery algorithm. Among such retrieval algorithms, PCGPA is the most popular \citep{delong1994pulse}, though it requires a full spectrogram that fulfills the Fourier relations and suffers from restrictions on the number of measurements. Another recent approach, Ptychographic FROG \citep{Sidorenko16}, offers improvement by handling any Fourier relations and partial measurements. However, PCGPA and Ptychographic FROG are able to reconstruct ultra-short pulses, but their performance deteriorates at low signal to noise ratio (SNR) to the extent that weak ultrashort pulses are currently considered non-recoverable. 

A different approach is to train a parametric model, e.g., a Deep Neural Network (DNN), to reconstruct the pulse from its measurement \citep{krumbugel1996direct}. However, previous attempts \citep{krumbugel1996direct} were restricted to primitive neural network implementations, which were limited to manual engineering of features, use of relatively simple and shallow neural networks and exploitation of small datasets. Their results were applied only to third harmonic generation (THG) FROG, which is not efficient in terms of measured power (thus highly not popular). Recently, DNNs have been used to automatically extract expressive features from data, leading to state-of-the-art learning results in image classification \cite{krizhevsky2012imagenet}, natural language processing \cite{sutskever2014sequence}, Go \cite{silver2016mastering}, visual diagnosis of skin cancer \cite{esteva2017dermatologist} and estimating the parameters of strong gravitational lensing systems \cite{grav_nature} to name a few. \\

Here, we propose and demonstrate, theoretically and experimentally, the reconstruction of ultrashort optical pulses by employing Deep Neural Networks (DNN). In particular, we train a convolutional neural network (CNN) to learn the inverse mapping of the FROG measurement function using supervised learning techniques. By doing that, we use convolutions not only to measure the pulse but also to reconstruct it. In addition, we observe that the FROG measurement function can be represented as an a-parametric DNN, and hence, it can be computed and differentiated using standard back-prop. Our simulations indicate that the concept works also for very weak pulses, which were thus far extremely hard (virtually impossible) to recover.  Finally, we show that we can combine unsupervised learning techniques on experimental data to achieve state-of-the-art results on measured pulses, and successfully overtake the existing reconstruction approach.

\section{Problem formulation}
\begin{figure}[h]
\centering
\includegraphics[width=\textwidth]{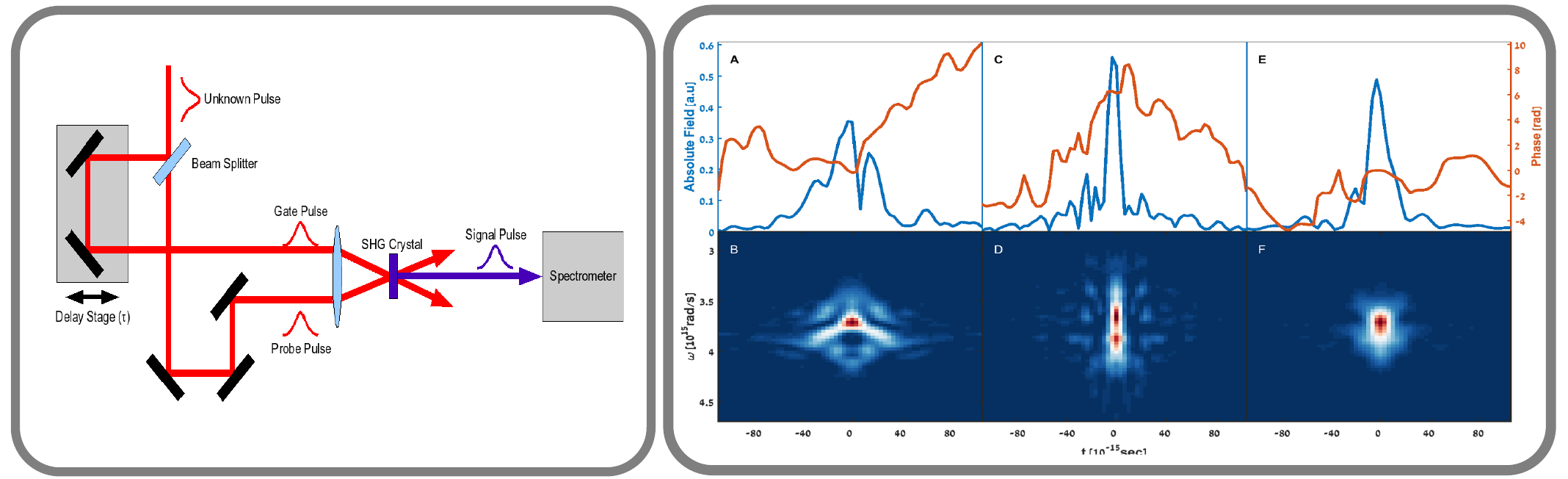}
\caption{Left: Experimental measurement setup. Right: Examples from the data set. Top: pulses amplitude (blue) and phase (red). Bottom: Corresponding FROG traces.}
\label{fig:setup}
\end{figure}

We will focus on the Second Harmonic Generation (SHG) FROG, a common example of a FROG system, where spectrograms of auto-correlations of an unknown complex pulse \(E(t)\) are measured to reconstruct it.  First, an SHG pulse is created from the unknown pulse \(E(t)\) and a shifted version of itself \(E(t-\tau)\) by projecting the two through an SHG crystal. This process results with an SHG pulse \(E_{SHG}(t)\) that is proportional to the product \(E(t)E(t-\tau)\). We then measure the Fourier amplitude of the SHG pulse using a discrete spectrometer, such that $I_{measured}(\omega_i,\tau)=|\mathcal{F}\{E(t)E(t-\tau)\}(\omega_i)|^2.$
We repeat this measurement for a set of time shifts and denote the outcome as the FROG trace:
$I_{measured}(\omega_i,\tau_j)=|\mathcal{F}\{E(t)E(t-\tau_j)\}(\omega_i)|^2.$ The reconstruction problem is defined by mapping the FROG trace to the pulse that created it $I_{measured}(\omega_i,\tau_j)\rightarrow\tilde{E}(t).$ Figure \ref{fig:setup} depicts the experimental system for generating FROG traces and examples of pulses and their FROG traces.

\textbf{Ambiguity removal.} Any FROG trace has a unique reconstructed pulse \(\tilde{E}(t)\) up to trivial ambiguities, e.g., constant phase shift, inversion with conjugate and translation (proof is based on the Fundamental Theorem of Algebra \citep{bendory2017uniqueness,trebino2000frequency}). This observation is essential, as it assures us that pulses with similar FROG traces are similar themselves. However, while these ambiguities are considered trivial and do not concern practitioners in physics, they can cause instabilities when training a DNN. Ambiguities are the regression equivalent of multiple labels in classification (see \citep{tsoumakas2006multi} for a survey), and are better to be avoided if possible. In practice, we noticed that by removing this ambiguity from our data set (mapping all of them into a singular group)  we could help the neural network perform much better.

\begin{figure}[H]
\centering
\includegraphics[width=\textwidth]{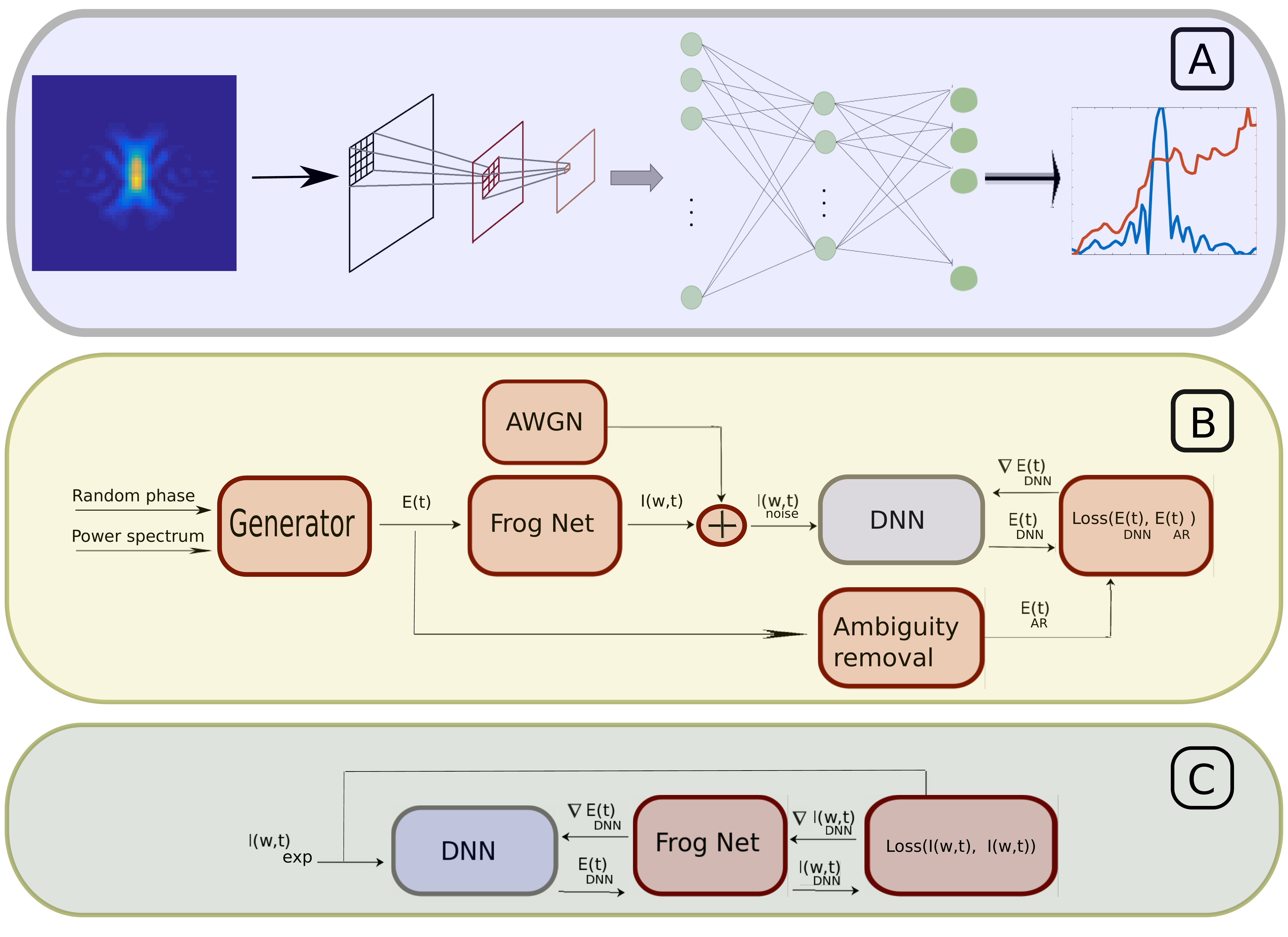}
\caption{\textbf{(A)} A DNN architecture for mapping FROG traces into complex pulses. \textbf{(B)} Supervised training procedure. A pulse is generated by a computer using a random phase and a power spectrum. The pulse is then split into two; one replica is used as a label for supervised training a DNN (after removing ambiguities from it). The second replica is passed through the FROG network to output its FROG trace, while White Gaussian Noise (AWGN) is added to it. The noisy FROG trace is then forward propagated in a DNN, compared with the first replica and produces a gradient. The gradient is back propagated through the DNN and is added to the weights through a Stochastic Gradient Descent (SGD) update. \textbf{(C)} Unsupervised training procedure. Similar to B, but now, the reconstructed pulse (the output of the DNN) is also forward propagated through the FROG net, such that the reconstructed FROG trace is compared with the measured one. The gradient is then computed and is backpedaled through the FROG net to update the weights of the DNN.}

\label{fig:methods}
\end{figure}

\section{Methods}
Our algorithm, denoted by DeepFROG, consists of two functions (represented as DNNs). The first function, denoted by FROG Net, is an a-parametric function representing the measurement system, i.e., given a pulse it produces a FROG trace. We represent this function using deep learning building blocks, such that it provides both evaluations (forward propagation) and gradients (backward propagation) of the function. We emphasize that this function remains constant, i.e., it is not learned or changed at any time. The second function is a parametric CNN, with weights $w$, that are optimized to be close to the inverse mapping of the Frog Net. We achieve this goal using Adam \citep{kingma2014adam}, a variant of Stochastic Gradient Descent (SGD), by minimizing the $l_1$ loss between the measurement and the reconstruction : 
\begin{equation}
w* = \underset{w}{\mbox{arg min}} \{ \mbox{ loss}(I,E) \} = \underset{w}{\mbox{arg min}} \{ \| \mbox{CNN}(I;w) - E \| _1 + \lambda \| \mbox{FROG Net}(\mbox{CNN}(I;w)) - I \| _1 \},
\label{eq:FROG}
\end{equation} where $\mbox{CNN}(I;w)$ denotes the output of the CNN function with weights $w$ given input $I$.

We experimented with three CNN \citep{lecun1990handwritten} architectures. Type 1 is a typical CNN, with three convolutional layers and two fully connected layers followed by ReLU activations. Model 2 uses multiple filters (with different filter sizes) at each convolution layer and denoted by Multires (inspired by \cite{szegedy2015going}), and type 3 is a densely connected network (DenseNet \cite{huang2016densely}). Overall, the DenseNet and Multires architectures performed the best, while Multires took a shorter time to reach good results. Due to space constraints, we only report results for the Multires model and leave architecture comparison for a more extended version of this work. \\

\section{Experiments}
\subsection{Sim2Sim}
\textbf{Setup.} We create a computer simulated data set in the following manner. We first generate a random spectral phase by multiplying a random complex flat-averaged vector and a Lorentzian envelope around the zero frequency. We then combine it with a Gaussian spectrum and transform it into a time-domain electromagnetic field using Fourier transform. This process ensures that the pulse is limited in time while having small, fast features. These pulses are then forward propagated in the FROG Net to output FROG traces. By repeating this process, we collected $60k $ training examples and $10k$ testing examples (examples that are only used for evaluation, and intentionally not for learning), each containing pairs of inputs (FROG traces, $I$) and labels (pulses, $E$). The entire learning process is described in details in Figure \ref{fig:methods}. \\

\begin{figure}[h]
\centering
\includegraphics[width=\textwidth]{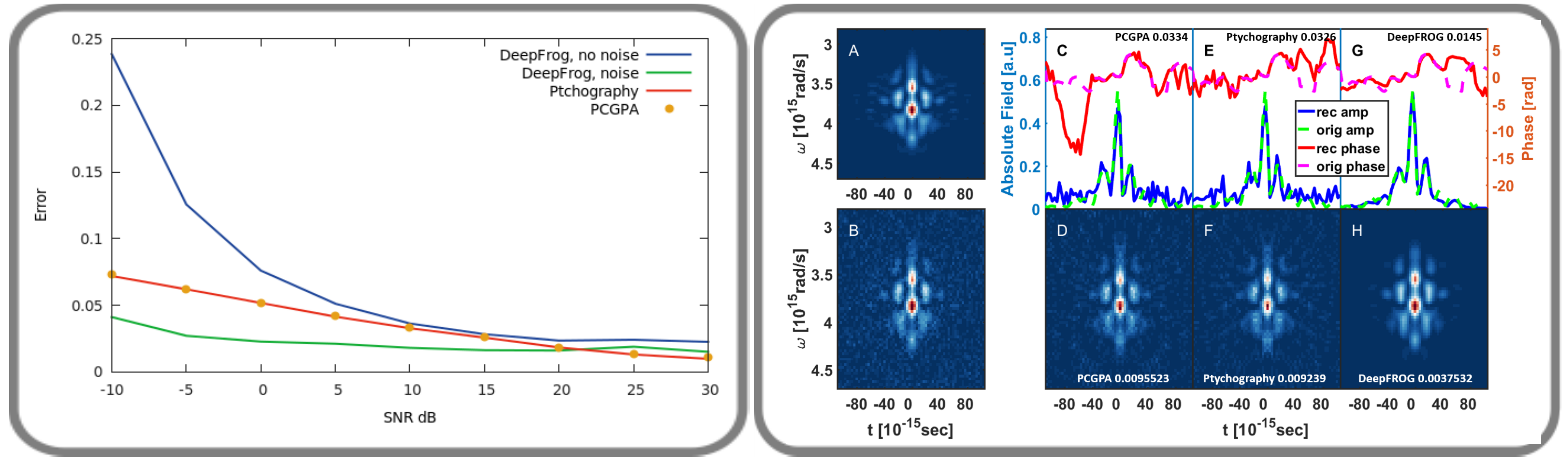}
\caption{Left: Reconstruction Error as a function of noise. DeepFROG was trained with (green) and without (blue) noise. Right: A) is the original spectogram and B)  is the result of adding 10dB Web. C), E)  and G) are the pulse reconstructions - amplitude and phase, compared to the original pulse and D), F), and H) are their spectrogram for PCGPA, Ptychograhpy and DeepFROG respectively.}
\label{fig:sim}
\end{figure}

\textbf{Results.} We now present numerical results on computer simulated data, both for training and evaluation. We trained two variants of the DeepFROG algorithm, one with injected noise (AWGN) in the range of $0$ to $30$ dB (sampled uniformly) and the second without injected noise. Figure \ref{fig:sim} presents the final performance of the different methods. On the left, we can see the performance of the networks, measured on unseen pulses with varying levels of noise. The DeepFROG variant that we trained with noise achieves lower reconstruction error than classical methods for SNR values below 20dB. On the right, we can see an example of a pulse from the simulated data set (with 10dB noise), along with its reconstructions. The DeepFROG algorithm produces better reconstructions with lower error compared to existing methods.

\subsection{Sim2Real}
\textbf{Setup.} To create ultra-short laser pulses in the lab, we use a mode-locked Ti-Sapphire laser and pass it through dispersive media to deform the pulse. The pulse is then being duplicated and later recombined when the replica of the pulse is shifted in time \(\tau\) concerning the other. At the recombination point, we place a nonlinear SHG crystal to create a second harmonic pulse which is proportional to the multiplication of the shifted pulses (the SHG phase matching was done by the anisotropy) and measures it with a spectrometer. This process is repeated $N$ times for different values of \(\tau\). We highlight that this process is the experimental equivalent of Equation \ref{eq:FROG}, represented by the FROG Net, as presented in Figure \ref{fig:setup} left.\\

\textbf{Method.} Our goal is to reconstruct the pulse, whose FROG trace is as close as possible to the measurement. For this purpose, we combine supervised learning on computer simulated data (similar to the previous section) with unsupervised learning (Figure \ref{fig:methods}, Bottom) on the measurement itself, i.e., we are using the gradients from the Frog Net, without knowing the reconstructed pulse itself. The unsupervised learning part allows us to use the structure of the problem (using the gradients of the FROG Net) and to make the solution specific for the experiment without using any knowledge on the pulse itself.\\

\begin{figure}[h]
\centering
\includegraphics[width=\textwidth]{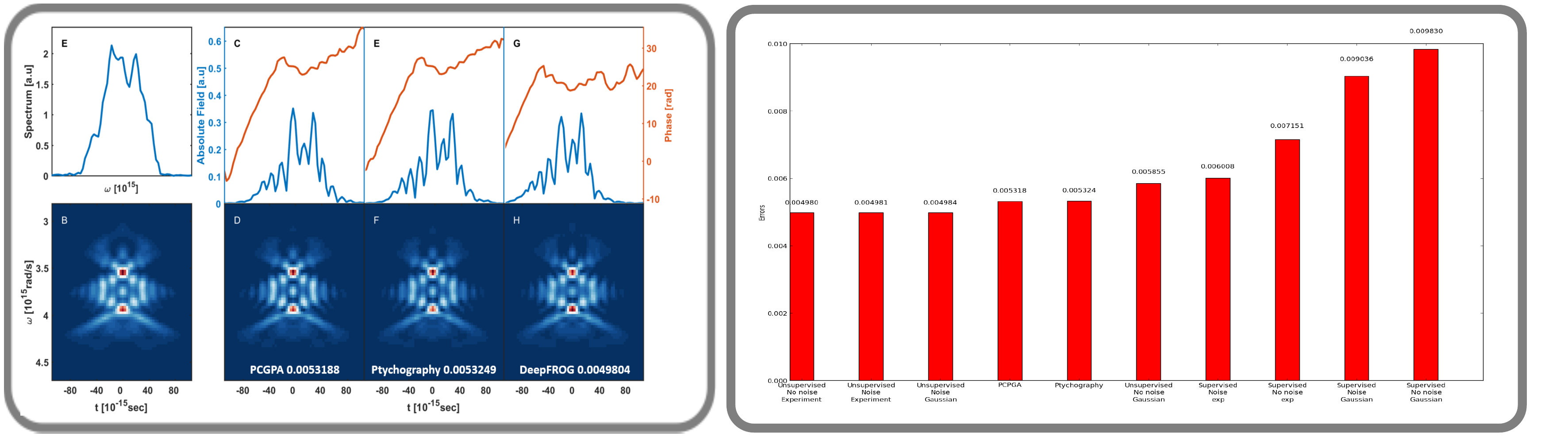}
\caption{Experimental results, Left: Reconstructed pulses and FROG traces for different algorithms. Right: Final reconstruction error of different reconstruction algorithms, measured by the mean absolute loss of the FROG trace.  }
\label{fig:exp_res}
\end{figure}

\textbf{Results.} Figure \ref{fig:exp_res} (Right), shows the reconstruction error of different methods. We compare state-of-the-art methods (Ptychography, PCGPA) with Deep Learning configurations. We present results for different deep learning configurations, e.g., unsupervised training, noise injection to training examples, and two types of power spectrum: a Gaussian and the true power spectrum of the pulse (used to create the supervised training data). Looking at Figure \ref{fig:exp_res} (Right), we can see that adding the unsupervised training was crucial to the success of our method. Also, having the true power spectrum didn't have a big effect on the results when using unsupervised training, but it did improve performance when we combined it with supervised learning. Overall, injecting noise during learning teaches the network to filter noise. Figure \ref{fig:exp_res} (Left) presents the reconstructed pulses and FROG traces of the experimental pulse, for the Ptychography, PCGPA and the best deep learning method. We can see that the deep learning method reached the lowest error. 

\section{Conclusions}
In this work, we presented a deep learning approach to reconstruct ultra-short laser pulses from their measured FROG traces. Our experiments suggest great potential of the approach to reducing reconstruction error for low SNR values. Also, we demonstrated the applicability of our method on experimental data measured in our lab.

\medskip
\small
\bibliography{paper_bib}

\begin{thebibliography}{}

\bibitem[\protect\citeauthoryear{Baltu{\v{s}}ka \bgroup et al\mbox.\egroup
  }{2003}]{baltuvska2003attosecond}
Baltu{\v{s}}ka, A.; Udem, T.; Uiberacker, M.; Hentschel, M.; Goulielmakis, E.;
  Gohle, C.; Holzwarth, R.; Yakovlev, V.; Scrinzi, A.; H{\"a}nsch, T.~W.;
  et~al.
\newblock 2003.
\newblock Attosecond control of electronic processes by intense light fields.
\newblock {\em Nature} 421(6923):611--615.

\bibitem[\protect\citeauthoryear{Bendory, Sidorenko, and
  Eldar}{2017}]{bendory2017uniqueness}
Bendory, T.; Sidorenko, P.; and Eldar, Y.~C.
\newblock 2017.
\newblock On the uniqueness of frog methods.
\newblock {\em IEEE Signal Processing Letters} 24(5):722--726.

\bibitem[\protect\citeauthoryear{Carquille, Da~Costa, and
  Froehly}{1975}]{carquille1975new}
Carquille, B.; Da~Costa, G.; and Froehly, C.
\newblock 1975.
\newblock New experimental methods for the analysis of light pulses and partial
  coherent laser beams.
\newblock {\em Japanese Journal of Applied Physics} 14(S1):17.

\bibitem[\protect\citeauthoryear{Delgado-Ru{\'\i}z \bgroup et al\mbox.\egroup
  }{2011}]{delgado2011femtosecond}
Delgado-Ru{\'\i}z, R.; Calvo-Guirado, J.; Moreno, P.; Guardia, J.;
  Gomez-Moreno, G.; Mate-S{\'a}nchez, J.; Ramirez-Fern{\'a}ndez, P.; and Chiva,
  F.
\newblock 2011.
\newblock Femtosecond laser microstructuring of zirconia dental implants.
\newblock {\em Journal of Biomedical Materials Research Part B: Applied
  Biomaterials} 96(1):91--100.

\bibitem[\protect\citeauthoryear{DeLong \bgroup et al\mbox.\egroup
  }{1994}]{delong1994pulse}
DeLong, K.~W.; Kohler, B.; Wilson, K.; Fittinghoff, D.~N.; and Trebino, R.
\newblock 1994.
\newblock Pulse retrieval in frequency-resolved optical gating based on the
  method of generalized projections.
\newblock {\em Optics letters} 19(24):2152--2154.

\bibitem[\protect\citeauthoryear{Esteva \bgroup et al\mbox.\egroup
  }{2017}]{esteva2017dermatologist}
Esteva, A.; Kuprel, B.; Novoa, R.~A.; Ko, J.; Swetter, S.~M.; Blau, H.~M.; and
  Thrun, S.
\newblock 2017.
\newblock Dermatologist-level classification of skin cancer with deep neural
  networks.
\newblock {\em Nature} 542(7639):115--118.

\bibitem[\protect\citeauthoryear{Gaumnitz \bgroup et al\mbox.\egroup
  }{2017}]{Gaumnitz17}
Gaumnitz, T.; Jain, A.; Pertot, Y.; Huppert, M.; Jordan, I.; Ardana-Lamas, F.;
  and W\"{o}rner, H.~J.
\newblock 2017.
\newblock Streaking of 43-attosecond soft-x-ray pulses generated by a passively
  cep-stable mid-infrared driver.
\newblock {\em Opt. Express} 25(22):27506--27518.

\bibitem[\protect\citeauthoryear{Hezaveh, Levasseur, and
  Marshall}{2017}]{grav_nature}
Hezaveh, Y.~D.; Levasseur, L.~P.; and Marshall, P.~J.
\newblock 2017.
\newblock Fast automated analysis of strong gravitational lenses with
  convolutional neural networks.
\newblock {\em Nature} 548(7669):555--557.

\bibitem[\protect\citeauthoryear{Huang \bgroup et al\mbox.\egroup
  }{2016}]{huang2016densely}
Huang, G.; Liu, Z.; Weinberger, K.~Q.; and van~der Maaten, L.
\newblock 2016.
\newblock Densely connected convolutional networks.
\newblock {\em arXiv preprint arXiv:1608.06993}.

\bibitem[\protect\citeauthoryear{Iaconis and
  Walmsley}{1998}]{iaconis1998spectral}
Iaconis, C., and Walmsley, I.~A.
\newblock 1998.
\newblock Spectral phase interferometry for direct electric-field
  reconstruction of ultrashort optical pulses.
\newblock {\em Optics letters} 23(10):792--794.

\bibitem[\protect\citeauthoryear{Kingma and Ba}{2014}]{kingma2014adam}
Kingma, D., and Ba, J.
\newblock 2014.
\newblock Adam: A method for stochastic optimization.
\newblock {\em arXiv preprint arXiv:1412.6980}.

\bibitem[\protect\citeauthoryear{Kop and Sprik}{1995}]{kop1995phase}
Kop, R.~H., and Sprik, R.
\newblock 1995.
\newblock Phase-sensitive interferometry with ultrashort optical pulses.
\newblock {\em Review of Scientific Instruments} 66(12):5459--5463.

\bibitem[\protect\citeauthoryear{Krizhevsky, Sutskever, and
  Hinton}{2012}]{krizhevsky2012imagenet}
Krizhevsky, A.; Sutskever, I.; and Hinton, G.~E.
\newblock 2012.
\newblock Imagenet classification with deep convolutional neural networks.
\newblock In {\em Advances in neural information processing systems},
  1097--1105.

\bibitem[\protect\citeauthoryear{Krumb{\"u}gel \bgroup et al\mbox.\egroup
  }{1996}]{krumbugel1996direct}
Krumb{\"u}gel, M.~A.; Ladera, C.~L.; DeLong, K.~W.; Fittinghoff, D.~N.;
  Sweetser, J.~N.; and Trebino, R.
\newblock 1996.
\newblock Direct ultrashort-pulse intensity and phase retrieval by
  frequency-resolved optical gating and a computational neural network.
\newblock {\em Optics letters} 21(2):143--145.

\bibitem[\protect\citeauthoryear{LeCun \bgroup et al\mbox.\egroup
  }{1990}]{lecun1990handwritten}
LeCun, Y.; Boser, B.~E.; Denker, J.~S.; Henderson, D.; Howard, R.~E.; Hubbard,
  W.~E.; and Jackel, L.~D.
\newblock 1990.
\newblock Handwritten digit recognition with a back-propagation network.
\newblock In {\em Advances in neural information processing systems},
  396--404.

\bibitem[\protect\citeauthoryear{Malinauskas \bgroup et al\mbox.\egroup
  }{2016}]{malinauskas2016ultrafast}
Malinauskas, M.; {\v{Z}}ukauskas, A.; Hasegawa, S.; Hayasaki, Y.; Mizeikis, V.;
  Buividas, R.; and Juodkazis, S.
\newblock 2016.
\newblock Ultrafast laser processing of materials: from science to industry.
\newblock {\em Light: Science \& Applications} 5(8):e16133.

\bibitem[\protect\citeauthoryear{Sidorenko \bgroup et al\mbox.\egroup
  }{2016}]{Sidorenko16}
Sidorenko, P.; Lahav, O.; Avnat, Z.; and Cohen, O.
\newblock 2016.
\newblock Ptychographic reconstruction algorithm for frequency-resolved optical
  gating: super-resolution and supreme robustness.
\newblock {\em Optica} 3(12):1320--1330.

\bibitem[\protect\citeauthoryear{Silver \bgroup et al\mbox.\egroup
  }{2016}]{silver2016mastering}
Silver, D.; Huang, A.; Maddison, C.~J.; Guez, A.; Sifre, L.; Van Den~Driessche,
  G.; Schrittwieser, J.; Antonoglou, I.; Panneershelvam, V.; Lanctot, M.;
  et~al.
\newblock 2016.
\newblock Mastering the game of go with deep neural networks and tree search.
\newblock {\em Nature} 529(7587):484--489.

\bibitem[\protect\citeauthoryear{Sutskever, Vinyals, and
  Le}{2014}]{sutskever2014sequence}
Sutskever, I.; Vinyals, O.; and Le, Q.~V.
\newblock 2014.
\newblock Sequence to sequence learning with neural networks.
\newblock In {\em Advances in neural information processing systems},
  3104--3112.

\bibitem[\protect\citeauthoryear{Szegedy \bgroup et al\mbox.\egroup
  }{2015}]{szegedy2015going}
Szegedy, C.; Liu, W.; Jia, Y.; Sermanet, P.; Reed, S.; Anguelov, D.; Erhan, D.;
  Vanhoucke, V.; and Rabinovich, A.
\newblock 2015.
\newblock Going deeper with convolutions.
\newblock In {\em Proceedings of the IEEE conference on computer vision and
  pattern recognition},  1--9.

\bibitem[\protect\citeauthoryear{Trebino \bgroup et al\mbox.\egroup
  }{1997}]{trebino1997measuring}
Trebino, R.; DeLong, K.~W.; Fittinghoff, D.~N.; Sweetser, J.~N.; Krumb{\"u}gel,
  M.~A.; Richman, B.~A.; and Kane, D.~J.
\newblock 1997.
\newblock Measuring ultrashort laser pulses in the time-frequency domain using
  frequency-resolved optical gating.
\newblock {\em Review of Scientific Instruments} 68(9):3277--3295.

\bibitem[\protect\citeauthoryear{Trebino}{2000}]{trebino2000frequency}
Trebino, R.
\newblock 2000.
\newblock {\em Frequency-Resolved Optical Gating: The Measurement of Ultrashort
  Laser Pulses: The Measurement of Ultrashort Laser Pulses}, volume~1.
\newblock Springer.

\bibitem[\protect\citeauthoryear{Tsoumakas and
  Katakis}{2006}]{tsoumakas2006multi}
Tsoumakas, G., and Katakis, I.
\newblock 2006.
\newblock Multi-label classification: An overview.
\newblock {\em International Journal of Data Warehousing and Mining, 3(3)}.

\bibitem[\protect\citeauthoryear{Zewail}{2000}]{zewail2000femtochemistry}
Zewail, A.~H.
\newblock 2000.
\newblock Femtochemistry: Atomic-scale dynamics of the chemical bond.
\newblock {\em The Journal of Physical Chemistry A} 104(24):5660--5694.

\end{thebibliography}
\bibliographystyle{aaai}

\end{document}